\documentclass{article}
\usepackage{amsmath}
\usepackage{amsthm}
\usepackage{clrscode}
\usepackage{tikz}

\title{Algorithms for Finding Dispensable Variables}
\date{}
\author{Mikol\'a\v s Janota \and Joao Marques-Silva \and Radu Grigore}

\begin{document}
\maketitle

\noindent \textsl{This short note reviews briefly three
algorithms for finding the set of dispensable variables of a
boolean formula. The presentation is light on proofs and heavy on
intuitions.}
\bigskip

It is sometimes desirable to find the set of variables that have
value~$0$ in all minimal models of a boolean formula~\cite{miko_dispensable}.
A \emph{minimal model} is one in which flipping any variable's
value from~$1$ to~$0$ leads to a non-model. All the models of the
function $a \oplus b$ are minimal ($01$ and $10$), and they are
also the minimal models of $a \lor b$, which has one non-minimal
model ($11$). For both these examples the set of dispensable
variables does not contain variable~$a$, nor variable~$b$.

\paragraph{Preliminary definitions.}
A \emph{literal} is a variable or the negation of a variable. A
\emph{clause} is a disjunction of literals, usually represented
as a set. A \emph{CNF} formula is a conjunction of clauses,
usually represented as a list. A \emph{model of a boolean
formula} is a map from variables to values that makes the
value of the formula~1.

\section{The MaxSAT approach}

A \emph{weighted MaxSAT solver} takes as input a CNF formula with
clauses $c_1,\dots,c_m$ and positive weights $w_1,\dots,w_m$
associated with each clause. The output is a model that
maximizes $\sum_i w_i c_i$.

A weighted MaxSAT solver can be used to find a \emph{cardinality
minimum model}, a model that has as few variables with value~$1$
as possible. Suppose the original clauses are $c_1,\dots,c_m$
and the variables are $v_1,\dots,v_n$. We add the \emph{clauses}
$\lnot v_1,\dots,\lnot v_n$, each with weight~$1$. We give to
each of the original clauses weight~$n+1$. The weighted MaxSAT 
problem is now\footnote{In fact, a non-weighted MaxSAT solver
that knows about \emph{hard} and \emph{soft} clauses is enough.}:

\begin{align}
\textrm{clauses: } &&c_1,\dots,&c_m,&\lnot v_1,\dots,&\lnot v_n &\\
\textrm{weights: } &&n+1,\dots,&n+1,&1,\dots,&1 & 
\end{align}

\noindent It is easy to see that the weighted MaxSAT solver will
satisfy $c_1,\dots,c_m$ when it is possible, and will choose
as many values of~$0$ for variables as possible.

\paragraph{A general approach.}
A cardinality minimum model is also a minimal model. (The converse
is false.) Once we can find \emph{one} minimal model we can generate
\emph{all} minimal models using the following algorithm:

\begin{codebox}
\Procname{$\proc{Generate-Minimal-Models}(f)$}
\li $\mu \gets \proc{Minimal-Model}(f)$
\li \While $\mu \ne \const{nil}$
\li \Do
      $\proc{Visit}(\mu)$
\li   $f \gets f \land \bigvee_{\mu(v)} \lnot v$
\li   $\mu \gets \proc{Minimal-Model}(f)$
    \End
\end{codebox}

\noindent Note that the number of minimal models may be exponential
in the number of variables and in the number of clauses:

\begin{align}
\bigwedge_{1\le k\le n} (v_{2k-1} \oplus v_{2k})
  &= (v_1 \oplus v_2) \land (v_3 \oplus v_4) \land \cdots \\
  &= (v_1 \lor v_2) \land (\lnot v_1 \lor \lnot v_2) \land
     (v_3 \lor v_4) \land (\lnot v_3 \lor \lnot v_4) \cdots
\end{align}

We are now investigating an approach that exploits the inner
workings of a MaxSAT solver. In particular, some MaxSAT solvers
use a \emph{bound} on the solution value, and that evolves
predictably when the formula is modified as in the previous
algorithm.

\section{The BDD approaches}

\emph{Binary decision diagrams} (BDD) are an
alternative to CNF for representing boolean formulas. The
function $\lnot v_0 \lor v_1$ and the function $\lnot v_1
\lor v_0$ have the following BDDs:

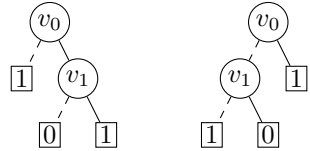
\begin{center}
\begin{tikzpicture}[scale=.5,every node/.style={draw,circle,inner sep=.5mm,solid}]
\node {$v_0$}
  child[dashed] { node[rectangle] {1} }
  child { node {$v_1$}
    child[dashed] { node[rectangle] {0} }
    child { node[rectangle] {1} }
  };
\end{tikzpicture}
\hskip 1cm
\begin{tikzpicture}[scale=.5,every node/.style={draw,circle,inner sep=.5mm,solid}]
\node {$v_0$}
  child[dashed] { node {$v_1$}
    child[dashed] { node[rectangle] {1} }
    child[solid] { node[rectangle] {0} }
  }
  child { node[rectangle] {1} };
\end{tikzpicture}
\end{center}

\noindent Even though the two functions are essentially the same,
the BDDs are different because one constraint of BDDs is to have
variables ordered on all paths from the root to a leaf. BDDs are
directed acyclic graphs. To evaluate the formula for a given
assignment of values to variables we start from the root and
at each node look at the value of the variable that labels the
node: If it is $0$ then we take the \emph{low} branch; if it is
$1$ we take the \emph{high} branch. The value of the function
is given by the leaf that is reached by this process. Another
constraint on BDDs is that they have no redundant node: There
is no node whose low and high branches point to the same place
and there are no two nodes that have the same label and their
respective branches point to the same place. (In particular, this
`no-redundancy' rule means we can't have two leafs with the same
value, but that would be difficult to draw.)

\paragraph{Reusing the general approach.}
Certain operations are particularly easy to carry out on BDDs.
For example we can find the \emph{lexicographically minimum model}
by starting at the root and always taking the low branch unless it
leads to~$0$. A lexicographically minimum model is also a minimal
model. (The converse is not true.) Therefore we can use the same
approach as before and implement the procedure $\proc{Minimal-Model}$
using BDDs. Preliminary experiments show that MaxSAT solvers tend
to work better in this context.

\paragraph{A BDD-specific solution.}
With BDDs we can:
\begin{enumerate}
\item Build a BDD that represents all \emph{minimal} models.
\item Extract the set of dispensable variables from this BDD.
\end{enumerate}

The function $\mathit{minimal}(f)$ gives the BDD whose all
models are the minimal models of the BDD~$f$.
\begin{equation}
\mathit{minimal}(v?h:l) = 
  v? (\mathit{minimal}(h)\land\mathit{monotone}(l)) : \mathit{minimal}(l)
\end{equation}

\noindent The notation $v?h:l$ denotes a BDD whose root node
is labeled by variable~$v$ and whose high and low branches are
pointing, respectively, to the BDDs $h$ and~$l$. A logical
operation applied to two BDDs, such as $\land$ above, is
understood to stand for the proper algorithm, which is outside
the scope of this short note.

The function $\mathit{monotone}(f)$ gives a BDD whose models are
all the models of~$f$ plus those that can be obtained by flipping
the value~$0$ into value~$1$ for some variables in a model. For
example, $\mathit{monotone}(a \oplus b)=a \lor b$. Interestingly,
in this case $\mathit{monotone}$ and $\mathit{minimal}$ are
inverses, since $\mathit{minimal}(a \lor b)=a \oplus b$.

\begin{equation}
\mathit{monotone}(v?h:l)=(v \land \mathit{monotone}(h))\lor\mathit{monotone}(l)
\end{equation}

Previously we did not discuss what procedure $\proc{Visit}$ does
to keep track of dispensable variables because it was obvious.
But it is worth mentioning how the set of dispensable variables
is obtained from $\mathit{minimal}(f)$. We can extract
the set of variables that have the value~$1$ in some model as
follows:

\begin{align}
\mathit{extract}(v?0:l)&=\mathit{extract}(l) \\
\mathit{extract}(v?h:l)&=\{v\}\cup\mathit{extract}(l)\cup\mathit{extract}(h)
\end{align}

\noindent The set of dispensable variables of a formula~$f$ 
is the complement of 
\begin{equation}
\mathit{extract}(\mathit{minimal}(f))
\end{equation}

\paragraph{A few words about efficiency.} Given BDDs~$f$ and~$g$
of sizes $m$ and~$n$ it takes $O(mn)$ time (and space) to compute
$f \circ g$ for any binary boolean operation~$\circ$. But
\emph{typically} it takes only time proportional to $m+n$. As a
result, a good (folklore to our knowledge) heuristic for going
from CNF to BDD is to construct a small BDD for each clause, put
them in a priority queue with the smallest at its root, and then
repetedly compute the binary operation $\land$ between the two
smallest BDDs. (The problem of minimizing the time is the same as
Huffman coding if time and space are both exactly $m+n$.)

Another interesting observation is that
$|\mathit{monotone}(f)|\le|f|$. Here $|f|$ denotes the size of
the BDD representing the function~$f$. (The result naturally
extends to the \emph{smallest} BDD under permutations of
variables.) To understand why this is so it is useful to think
of BDD nodes as being tagged with truth-tables~\cite{knuth_bdd}.
For example, the truth-table of $a \oplus b$ is $0110$ and it
labels the root of the corresponding BDD. The low branch points
to a node labeled by the first half $01$ and the high branch
points to $10$. Therefore the nodes in the BDD of $a \oplus b$
are $0110$, $01$, $10$, $0$, and $1$ for a total size of~$5$.
On the other hand the truth-table of $a \lor b$ is $0111$. The
nodes in this case are $0111$, $01$, $0$, and~$1$, for a total
size of~$4$. Notice that there is no node $11$. In fact there is
never a square node (of the form $aa$ for some~$a$) because of
the restriction that low and high branches are different. When
we compute the $\mathit{monotone}$ function the truth table $lh$
becomes $l(l\lor h)$, where $\lor$ is bitwise: This operation
(carried out recursively) may introduce square tables but may
never remove them. Qed.

We are now exploring the relation between
$|\mathit{minimal}(f)|$ and~$|f|$. In our experiments it is
almost always the case that $|\mathit{miminal}(f)|\le|f|$, but we know
this relation does not hold for $f=a \lor b$.

\bibliographystyle{plain}
\bibliography{minimal}

\end{document}